\documentclass[pra, 12pt, letter, floatfix]{revtex4-1}

\usepackage[version=3]{mhchem} 
\usepackage[T1]{fontenc}       

\usepackage{amsmath}
\usepackage{amsfonts}
\usepackage{amssymb}
\usepackage{graphicx}
\usepackage{dcolumn}
\usepackage{bm}
\usepackage{subcaption}
\usepackage{grffile}
\usepackage{verbatim}

\begin{document}

\title[Size consistent excited states] {Size consistent excited states via algorithmic transformations between variational principles}

\author{Jacqueline A. R. Shea$^{1}$}
\author{Eric Neuscamman$^{1,2,}$}
\email{eneuscamman@berkeley.edu}
\affiliation{$^1$Department of Chemistry, University of California, Berkeley, California 94720, USA\\
             $^2$Chemical Sciences Division, Lawrence Berkeley National Laboratory, Berkeley, California 94720, USA}

\date{\today}

\begin{abstract}
We demonstrate that a broad class of excited state variational
principles is not size consistent.
In light of this difficulty, we develop and test an approach
to excited state optimization
that transforms between variational principles in order
to achieve state selectivity,
size consistency, and compatibility with quantum Monte Carlo.
To complement our formal analysis, we provide numerical examples
that confirm these properties and demonstrate how they contribute
to a more black box approach to excited states in
quantum Monte Carlo.
\end{abstract}

\maketitle

\section{Introduction}
\label{sec:introduction}

In a large range of chemical and materials applications, including homogeneous catalysis,
liquid-phase light harvesting, and band gap engineering, it is essential that
the predictions made by theoretical methods retain their accuracy as the system size
is varied.
For example, a method whose accuracy was highly dependent on the number of solvent
molecules included in a simulation is less useful than a method whose accuracy is not.
Likewise, when performing solid-state calculations on a series of increasingly large
simulation cells in order to get a grip on finite-size effects \cite{shulenburger:2013:qmc_solids}, it is important that
a method's accuracy not be dependent on the system size, or else it will be difficult
to separate real finite size effects from other methodological artifacts.
At the very least, it is desirable that methods used in these contexts satisfy
size consistency, which demands that two separated systems produce the same results
whether they are simulated independently or together.

Famously, not all wave function approximations satisfy size consistency.
While the coupled cluster ansatz does, truncated configuration interaction
does not \cite{Bartlett:2007:cc_rev,MolElecStruc,Szabo-Ostland,Martin:1989:ccsd(t),Bartlett:1982:ccsd}.
Likewise, the single Slater determinant of standard Hartree-Fock theory is
size consistent, but symmetry-projected Hartree-Fock theory is not
\cite{Szabo-Ostland,scuseria:2011:pqt,Scuseria:2012:PHF}.
The antisymmetric geminal power (AGP)
\cite{POPLE:1953:agp,coleman1997agp,Scuseria:2002:HFB_optimization,Sorella:2003:jagp,Sorella:2007:weak_binding,Sorella:2009:rvb_molecules}
is not size consistent when used alone, but becomes size consistent
when paired with the right type of Jastrow factor (JAGP) \cite{Neuscamman:2012:sc_jagp,Neuscamman:2013:hilbert_jagp}.
Indeed, when designing new or improved wave function ansatzes, an important
theoretical test is to check whether or not size consistency is retained.

Like wave function ansatzes, variational principles come in both size consistent
and size inconsistent varieties.
The most famous and widely used variational principle, the ground state energy,
is of course size consistent, but others, including some used for the direct
optimization of excited states \cite{Mesmer:1970:esvp,Zhao:2016:dir_tar}, are not.
Indeed, even when paired with a size consistent wave function (e.g.\ one that product factorizes)
such size inconsistent variational principles can lead to size inconsistent results.
Thus, when designing variational principles and methods based on them, it is
important to consider the consequences that different choices will have on
size consistency.

Of course, many other properties, not least of which is affordability, must be
considered when designing principles and algorithms for use in the optimization
of wave functions.
For example, one can imagine incorporating higher powers of the Hamiltonian
operator when constructing a new variational principle, although in practice
it is quite rare to see powers higher than two for the simple reason that
higher Hamiltonian powers tend to lead to higher evaluation costs.
As accurate electronic structure methods are already quite computationally
intensive, it is not appealing to raise costs further.

Unfortunately, there is a strong formal problem that arises for excited
states when limiting the functional form of a variational principle to
include only the first and second power of the Hamiltonian.
As we prove in this paper, such variational principles cannot simultaneously
target an individual excited state and remain size consistent.
In light of this challenge, we advocate that in practice a wave function
optimization method intended for use with excited states may be best
served by amalgamating multiple variational principles.
For example, as was achieved recently by the $\sigma$-SCF method, \cite{VanVoorhis:2016:sigmaSCF}
an optimizer might begin by minimizing a size inconsistent but state specific
variational principle, but upon approaching convergence gradually
transition to minimizing a size consistent but state nonspecific
variational principle.
The idea is for the first variational principle to get the optimization
close enough to the desired eigenstate so that the lack of
state specificity in the final variational principle is no longer an issue.

Following our formal proof, we will present one such amalgamation
that works in the context of wave function optimization via
quantum Monte Carlo (QMC). \cite{FouMitNeeRaj-RMP-01}
Crucially, QMC can work with many excited state variational
principles for a cost similar to its ground state cost \cite{Zhao:2016:dir_tar},
and ground state QMC can reach scales up to hundreds
of atoms thanks to its low scaling and easy parallelization
\cite{Kim:2012:qmcpack_scaling,shulenburger:2013:qmc_solids}.
The realization of a state specific and size consistent
excited state optimizer in QMC thus marks an important step
towards achieving more reliable predictions of
excited states and spectral properties in complicated
molecules and materials.

\section{Theory}
\label{sec:theory}

\subsection{Variational Principles}
\label{sec::vp}

For the purposes of this paper, let us define a state selective variational principle as a smooth function of a wave
function ansatz's variables with the following property: if the ansatz is capable of exactly describing the individual
Hamiltonian eigenstate of interest, then the function will have its unique global minimum at the variable values
corresponding to that exact eigenstate.
If the state being targeted is the ground state, as occurs for the function
\begin{align}
\label{eqn:e_func}
E(\Psi) = \langle\hat{H}\rangle = \frac{ \langle\Psi|\hat{H}|\Psi\rangle }{ \langle\Psi|\Psi\rangle },
\end{align}
then we will call the function a ground state variational principle.
An excited state variational principle is, therefore, a state selective variational principle
for which the global minimum corresponds to an excited state.

Note that the energy variance
\begin{align}
\label{eqn:sigma2_func}
\sigma^2(\Psi) = \frac{ \langle\Psi|(\hat{H}-E)^2|\Psi\rangle }{ \langle\Psi|\Psi\rangle } = \langle\hat{H}^2\rangle - \langle\hat{H}\rangle^2,
\end{align}
can be employed as a variational principle,
\cite{Messmer:1969:esvp,Guzhavina:JSC:1982,Umrigar:1988:var_min,MCM_quant_chem,Umrigar:2005:e_v_opt,McClean:2016:var_quant_class}
but that it is not state selective, as its global minimum is not unique.
Indeed, any Hamiltonian eigenstate gives the equally low value of $\sigma^2=0$.
As we will discuss further in Section \ref{sec:compare_to_variance}, this lack of state selectivity can make optimization to the desired eigenstate
more difficult than one would prefer.

To be practical, a variational principle must be paired with an efficient method for its evaluation and minimization.
This requirement more or less explains why the energy-based ground state variational principle has been
more successful than variational principles for excited states.
Note that the functional form of $E$ requires an expectation value of only the first
power of the Hamiltonian, in contrast to $\sigma^2$ whose evaluation requires expectation values of
both $\hat{H}$ and $\hat{H}^2$, the latter of which is in most circumstances more computationally demanding.
Indeed, the construction of excited state variational principles that work by measuring a wave function's
``energetic distance'' from a desired position $\omega$ in the spectrum, such as
\cite{weinstein:1934:mod_ritz,macdonald:1934:mod_ritz,Mesmer:1970:esvp,VanVoorhis:2016:sigmaSCF}
\begin{align}
\label{eqn:theta_func}
W(\Psi) = \frac{ \langle\Psi|(\omega - \hat{H})^2|\Psi\rangle }{ \langle\Psi|\Psi\rangle }
= (\omega-E)^2 + \sigma^2
\end{align}
or \cite{Zhao:2016:dir_tar}
\begin{align}
\label{eqn:omega_func}
\Omega(\Psi) = \frac{ \langle\Psi|(\omega - \hat{H})|\Psi\rangle }{ \langle \Psi|(\omega - \hat{H})^2|\Psi\rangle }
= \frac{\omega-E}{(\omega-E)^2+\sigma^2},
\end{align}
tend to also require $\hat{H}^2$, because computing a distance typically involves taking a square.
In this respect, variational Monte Carlo (VMC) offers the advantage that $\hat{H}^2$ expectation values can
be evaluated via Monte Carlo integration \cite{Umrigar:2005:e_v_opt,Zhao:2016:dir_tar} of the integral
\begin{align}
\label{eqn:sigma2_qmc}
\langle\hat{H}^2\rangle =
\frac{\langle\Psi|\hat{H}^2|\Psi\rangle}{\langle\Psi|\Psi\rangle} = 
\int \hspace{0.5mm}
\frac{|\langle\Psi|\vec{r}\rangle|^2}{\langle\Psi|\Psi\rangle}
\left|\frac{\langle\vec{r}|\hat{H}|\Psi\rangle}{\langle\vec{r}|\Psi\rangle}\right|^2
d\vec{r}.
\end{align}
This approach avoids having to explicitly square the Hamiltonian operator and is thus similar in
difficulty to a VMC evaluation of the energy $E$.
In principle, variational principles that depend on cubic or higher powers of $\hat{H}$ could be
constructed, but these are likely to be even less practical, and so we will for this study limit
our attention to quadratic and lower powers of $\hat{H}$.

\subsection{The $V_{1,2}$ set}
\label{sec::v12_set}

To be size consistent, a method must predict the same total energy for completely separate
subsystems $A$ and $B$ whether treating them separately or together.
This occurs, for example, when the ground state variational principle is paired with a
product factorizable ansatz,
\begin{align}
\label{eqn:prod_fact}
\Psi_{AB} = \Psi_A \hspace{0.7mm} \Psi_B.
\end{align}
In this section, we define a broad class of state selective excited state variational principles.
In the next section, we will show that these fail to satisfy size consistency even when the ansatz is product factorizable.

To begin, let us define $V_{1,2}$ as the set of all state selective variational principles
that have the following three properties.
First, in the interest of affordable evaluation, we require the functional form of
any $\Gamma\in V_{1,2}$ to depend on the wave function variables only through
the expectation values $\langle\hat{H}\rangle$ and $\langle\hat{H}^2\rangle$.
Using Eqs.\ (\ref{eqn:e_func}) and (\ref{eqn:sigma2_func}), we see that this is the same
as requiring that $\Gamma$ depend on the wave function only through $E$ and $\sigma^2$,
\begin{align}
\label{eqn:vp_dependence}
\Gamma(\Psi) = \Gamma(E(\Psi), \sigma^2(\Psi)).
\end{align}
Second, we require that $\Gamma$ have a unique global minimum corresponding to a particular
interior and nondegenerate eigenstate of $\hat{H}$, thus limiting the analysis
to nondegenerate excited states and excluding both $\Gamma=E$ and $\Gamma=\sigma^2$
as possibilities.
Finally, we require $\Gamma$ to be real analytic (i.e.\ real valued and equal to its Taylor series)
in a contiguous, open region around the global minimum $(E,\sigma^2)=(E_t,0)$,
where $E_t$ is the energy of the targeted eigenstate.
Upon close inspection, one finds that both
$\Omega\in V_{1,2}$ and $W\in V_{1,2}$ when $\omega$ is close to but below $E_t$.

Note that we should not expect results to be size consistent if the wave function ansatz
cannot be product factorized when dealing with isolated systems.
Unless stated otherwise, we therefore assume that we are working with two completely
separate subsystems $A$ and $B$ and that our overall ansatz
can be written as a product of separate ansatzes for the two subsystems.
In this case, both the energy and variance will be additive:
\begin{align}
\label{eqn:addative_energy}
E_{AB} &= E_A + E_B, \\
\label{eqn:addative_variance}
\sigma^2_{AB} &= \sigma^2_A + \sigma^2_B,
\end{align}
where $E_{AB}=E(\Psi_{AB})$ is the energy of the system when evaluated as a combined whole
and $E_A=E(\Psi_A)$ is the energy of $A$ when treated alone.

\subsection{Proof of no size consistency}
\label{sec::proof_no_sc}

We now proceed to show that any variational principle within $V_{1,2}$
is not size consistent.
First, note some general properties that the Taylor series of $\Gamma$,
\begin{align}
\label{eqn:gamma_taylor}
\Gamma(E,\sigma^2) = \sum_{m=0}^\infty \sum_{n=0}^\infty a_{mn} (\sigma^2)^m (E-E_t)^n,
\end{align}
must satisfy if it is to be a member of $V_{1,2}$.
To start, we note that there must be a nonzero coefficient among the terms with $m=0$
and $n>1$.
If there were not, then either $\Gamma$ would not be state specific or it would not target
an interior eigenstate.
Similarly, there must be a nonzero coefficient among the terms with $n=0$ and $m>0$,
or else any state with $E_t$ as its energy expectation value would give the same value
for $\Gamma$ as the targeted eigenstate.
Finally, for ease of analysis and without loss of generality, we will set $a_{00}=0$,
as this does not alter the nature of the global minimum.
With these restrictions and defining $\Delta=E-E_t$, we can write the Taylor series as
\begin{align}
\label{eqn:gamma_taylor_3}
\Gamma = \sum_{m=p}^\infty                   a_{m0} (\sigma^2)^m
       + \sum_{m=q}^\infty \sum_{n=r}^\infty a_{mn} (\sigma^2)^m \Delta^n
       +                   \sum_{n=s}^\infty a_{0n}              \Delta^n
\end{align}
in which $p$, $q$ $r$, and $s$ are positive integers,
$a_{p0}\ne 0$, $a_{0s}\ne 0$, and it is understood that there exist
$1<\tilde{s}<\infty$ and $0<\tilde{p}<\infty$ such that $a_{0\tilde{s}}\ne 0$
and $a_{\tilde{p}0}\ne 0$.
Note that it is possible for all the elements in the middle sum to have zero coefficients, but
they can in general be nonzero.
In the latter case, $a_{qr}\ne 0$.

With these qualities of $\Gamma$ in mind, consider the stationary condition
\begin{align}
\label{eqn:stationary_wrt_x}
 0 = 
     \frac{\partial \Gamma}{\partial \Delta} \frac{\partial E}{\partial x}
   + \frac{\partial \Gamma}{\partial \sigma^2} \frac{\partial \sigma^2}{\partial x}
\end{align}
for minimizing $\Gamma$ when system $A$ is treated alone with an ansatz depending on a single variable $x$.
For any choice of $\Gamma\in V_{1,2}$, we show in Appendix A that there exist system/ansatz
pairs for which neither $E$ nor $\sigma^2$ are stationary at the global minimum of $\Gamma$.
Choosing this type of system/ansatz pair for system $A$ and defining the analytic functions
\begin{align}
\label{eqn:mu_expansion}
\mu(\Delta,\sigma^2) \equiv \frac{\partial \Gamma}{\partial \Delta}
= \sum_{m=q}^\infty \sum_{n=r}^\infty \hspace{0.5mm} n \hspace{0.5mm} a_{mn} (\sigma^2)^m (\Delta)^{n-1}
+                   \sum_{n=s}^\infty \hspace{0.5mm} n \hspace{0.5mm} a_{0n}              (\Delta)^{n-1}
\end{align}
and
\begin{align}
\label{eqn:nu_expansion}
\nu(\Delta,\sigma^2) \equiv \frac{\partial \Gamma}{\partial \sigma^2}
= \sum_{m=p}^\infty                   \hspace{0.5mm} m \hspace{0.5mm} a_{m0} (\sigma^2)^{m-1}
+ \sum_{m=q}^\infty \sum_{n=r}^\infty \hspace{0.5mm} m \hspace{0.5mm} a_{mn} (\sigma^2)^{m-1} (\Delta)^n
\end{align}
we can rewrite the stationary condition as
\begin{align}
\label{eqn:mu_nu_stat_a}
0 =   \mu(\Delta,\sigma^2) \frac{\partial        E}{\partial x}
    + \nu(\Delta,\sigma^2) \frac{\partial \sigma^2}{\partial x}.
\end{align}
When system $A$ is alone, this condition is satisfied for some $x=x_A$ at which $\Delta=\Delta_A$,
$\sigma^2=\sigma^2_A$, $\partial E/\partial x \ne 0$ and $\partial \sigma^2/\partial x \ne 0$.
Now imagine if we added a system $Q$ that is completely separated from system $A$ such that
$\hat{H}=\hat{H}_A+\hat{H}_Q$.
Choosing the overall wave function ansatz to be a product of the ansatzes from $A$ and $Q$
such that Eqs.\ (\ref{eqn:addative_energy}) and (\ref{eqn:addative_variance}) apply,
we see that $\Delta$ and $\sigma^2$ will be the only parts of Eq.\ (\ref{eqn:mu_nu_stat_a})
affected by the addition of $Q$ so long as $x$ is held fixed at $x=x_A$.
Crucially, note that $\partial E/\partial x$ and $\partial \sigma^2/\partial x$ are not affected.

We now separate $V_{1,2}$ into two subsets and show
that size consistency is violated in both.
First, take the subset in which the middle sum of Eq.\ (\ref{eqn:gamma_taylor_3}) is absent,
$p=1$, and $a_{m0}=0$ for $m>1$.
In this case, the right hand side of Eq.\ (\ref{eqn:mu_nu_stat_a}) will not be a function of $\sigma^2$.
If we were to hold $x=x_A$ fixed, then it would
be a nonconstant and analytic function of $\Delta$, and thus by the principle of permanence
its root at $\Delta=\Delta_A$ would be isolated.
This implies that upon adding a system $Q$ so that $\Delta\rightarrow\Delta_A+\Delta_Q$,
the stationary condition would for small but nonzero $\Delta_Q$ no longer 
be satisfied unless the value of $x$ were adjusted.
Thus, in this subset of $V_{1,2}$, the addition of system $Q$ would change the optimal
wave function in system $A$, despite the overall wave function product factorizing
and the two subsystems not interacting.

\begin{figure}[t]
\includegraphics[width=3.33in]{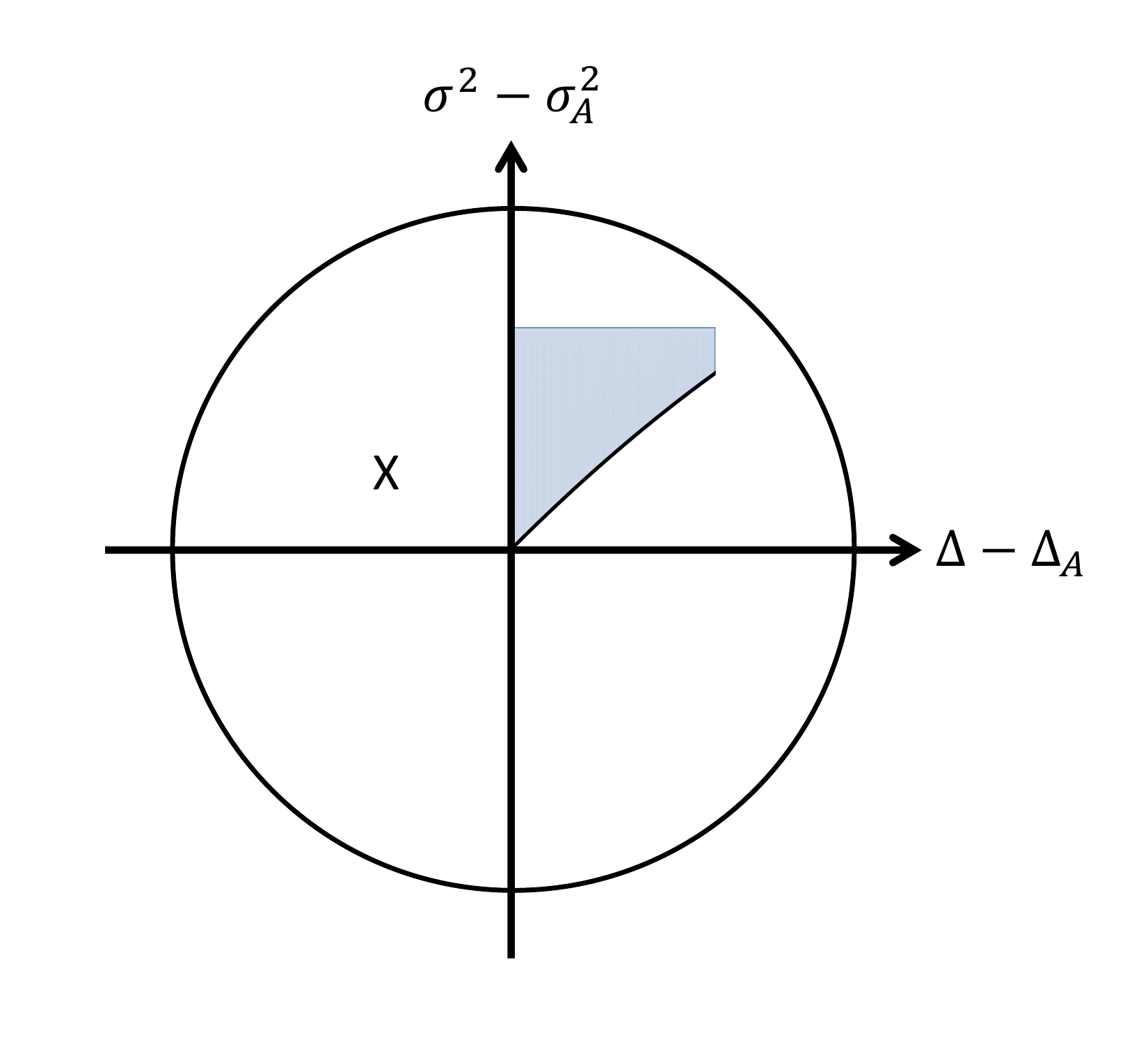}
\caption{A schematic showing values of $\Delta$ and $\sigma^2$ that can be reached by
         adding systems of types $B$ and $C$.  The line shows values accessible
         by the addition of systems of type $B$.  The shaded region shows values
         accessible by the addition of systems of types $B$ and $C$ together.
         The circle and the X give examples for $\Gamma$'s analytic region
         and global minimum, respectively.}
\label{fig:delta_sigma}
\end{figure}

Second, take the subset that contains all $\Gamma\in V_{1,2}$ not in the previous subset.
In this case, if we were to hold $x=x_A$
fixed, the right hand side of Eq.\ (\ref{eqn:mu_nu_stat_a}) will be an analytic function that depends
on both $\Delta$ and $\sigma^2$.
By now adding one subsystem of type $B$ and one of type $C$ (see Appendix B) such that none of
the three subsystems interact and the overall wave function is a product of the three
subsystem wave functions, we will have
\begin{align}
\Delta &= \Delta_A + \Delta_B + \Delta_C \\
\sigma^2 &= \sigma^2_A + \sigma^2_B + \sigma^2_C
\end{align}
in which
\begin{align}
  \Delta_B &= \frac{\alpha^2}{1+\alpha^2}, \\
\sigma^2_B &= \frac{\alpha^2}{(1+\alpha^2)^2}, \\
  \Delta_C &= 0, \\
\sigma^2_C &= \frac{\beta^2}{1+\beta^2},
\end{align}
where $\alpha$ and $\beta$ are real numbers.
By choosing different systems $B$ and $C$, we may vary $\alpha$ and $\beta$
to map out a contiguous two-dimensional patch within the region on which $\Gamma$ is analytic,
as shown in Figure \ref{fig:delta_sigma}.
As we can choose the system/ansatz pair in $A$ such that its stationary point $(\Delta_A,\sigma^2_A)$
is arbitrarily close to the global minimum, we may assume without loss of generality that
this patch is inside the region within which $\Gamma$ is analytic.
If the stationary condition in Eq.\ (\ref{eqn:mu_nu_stat_a}) were satisfied at all points in the patch, then by
repeated use of the principle of permanence, we see that it would also be satisfied
at all points in an open region encompassing the global minimum.
As this would violate our assumption of a unique global minimum, 
we must conclude that at the vast majority of points in the mapped-out patch,
i.e.\ for most choices of systems $B$ and $C$ with small $\alpha$ and $\beta$,
Eq.\ (\ref{eqn:mu_nu_stat_a}) will not be satisfied when $x=x_A$.
In other words, the addition of these completely separate subsystems changes the optimal
wave function in system $A$.
As this will in turn change the energy, we see that size consistency is violated.

To summarize, we have found that for any $\Gamma\in V_{1,2}$, it is possible to
construct a product separable ansatz for completely separate subsystems in
such a way that the optimal wave function on one subsystem is changed by the presence
of other subsystems.
As a result, the total energy will be different if we treat the systems separately instead of together.
We must thereby conclude that there are no size consistent variational principles in $V_{1,2}$.

\subsection{Transformations between variational principles}
\label{sec::update_omega_sec}

While individual members of $V_{1,2}$ are not size consistent, it is nonetheless possible to employ them as part
of an overall optimization scheme that is both state selective and size consistent.
As was achieved for Slater determinants in the $\sigma$-SCF method \cite{VanVoorhis:2016:sigmaSCF}, the general
strategy is to begin the optimization with a state selective variational principle in order to ensure the
correct state is targeted.
Once the wave function was ``close'' to the desired state, $\sigma$-SCF prescribed a transition to
state nonspecific variance minimization, which, among other benefits, ensures size consistency.
Here, we present an evolution of this general strategy that both makes it compatible with VMC and guarantees that
state selectivity is maintained throughout the optimization, even in the final stage in which
size consistency is achieved through variance minimization.

The key to our strategy is to recognize that special choices for $\omega$ can make either $W$ or $\Omega$ (and likely many other members of $V_{1,2}$)
become akin to variance minimization.
For example, one sees that
\begin{align}
\label{eqn:w_to_variance}
W(\Psi) \Big|_{\omega \rightarrow E} = \sigma^2
\end{align}
and
\begin{align}
\label{eqn:omega_to_variance}
\Omega(\Psi) \Big|_{\omega \rightarrow E-\sigma} = -\frac{1}{2\sigma}.
\end{align}
If, in the final stage of the optimization, we ensure that $\omega$ is chosen appropriately and self consistently, then minimizing $W$ or
$\Omega$ for a particular state will produce the same result as if variance minimization had been achieved for that state.
Crucially, we adopt a strategy in which $\Psi$ and $\omega$ are updated \textit{separately} in a ``tick-tock'' fashion, which ensures that
$\omega$ is fixed during an update step for $\Psi$.
This choice guarantees that the desired state is targeted, as it remains the global minimum
of $W$ or $\Omega$ during the $\Psi$ update step.
If we instead simply switched to variance minimization, we would in general have to rely on the state in question being a stable local minimum
of the variance, which does not offer the same convergence guarantees as a state selective approach in which the desired state is the global minimum.
Between each $\Psi$ update, we adjust $\omega$ to its special value (e.g.\ $E$ or $E-\sigma$) so that at convergence the result is
equivalent to variance minimization and thus size consistent.
Although we have chosen to test this strategy using $\Omega$ as the variational principle and the VMC linear method
\cite{Nightingale:2001:linear_method,UmrTouFilSorHen-PRL-07,TouUmr-JCP-07,TouUmr-JCP-08,Zhao:2016:dir_tar,Zhao:2017:blm}
as the wave function update method,
we expect it to be effective for other variational principles and updated methods as well.

\begin{figure}[h]
\includegraphics[width=3.33in]{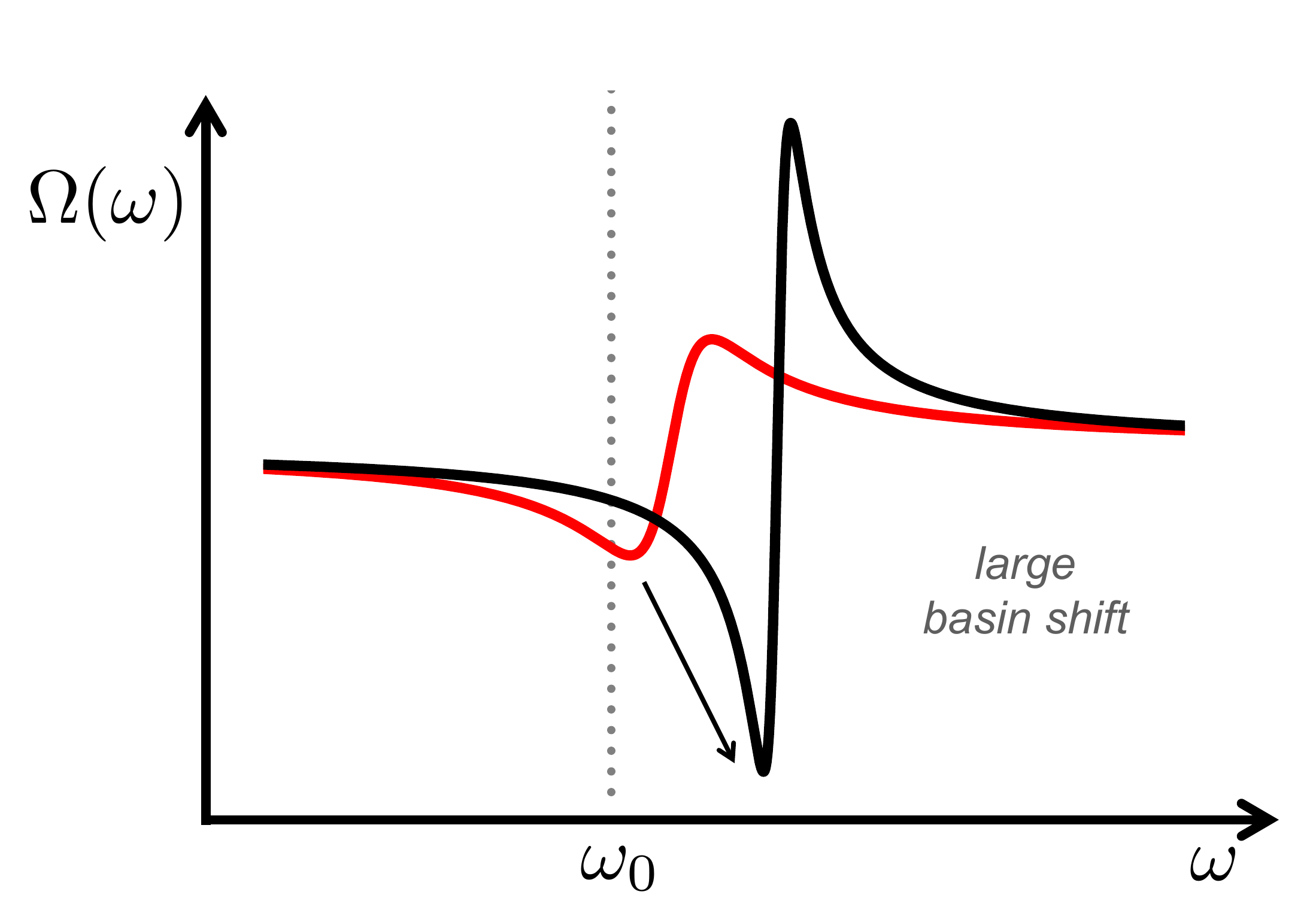}
\caption{Example of a large basin shift for $\Omega$ that may occur if we skip the transitional-$\omega$ stage of the optimization.
        }
\label{fig:large_basin_shift}
\end{figure}

\begin{figure}[h]
\includegraphics[width=3.33in]{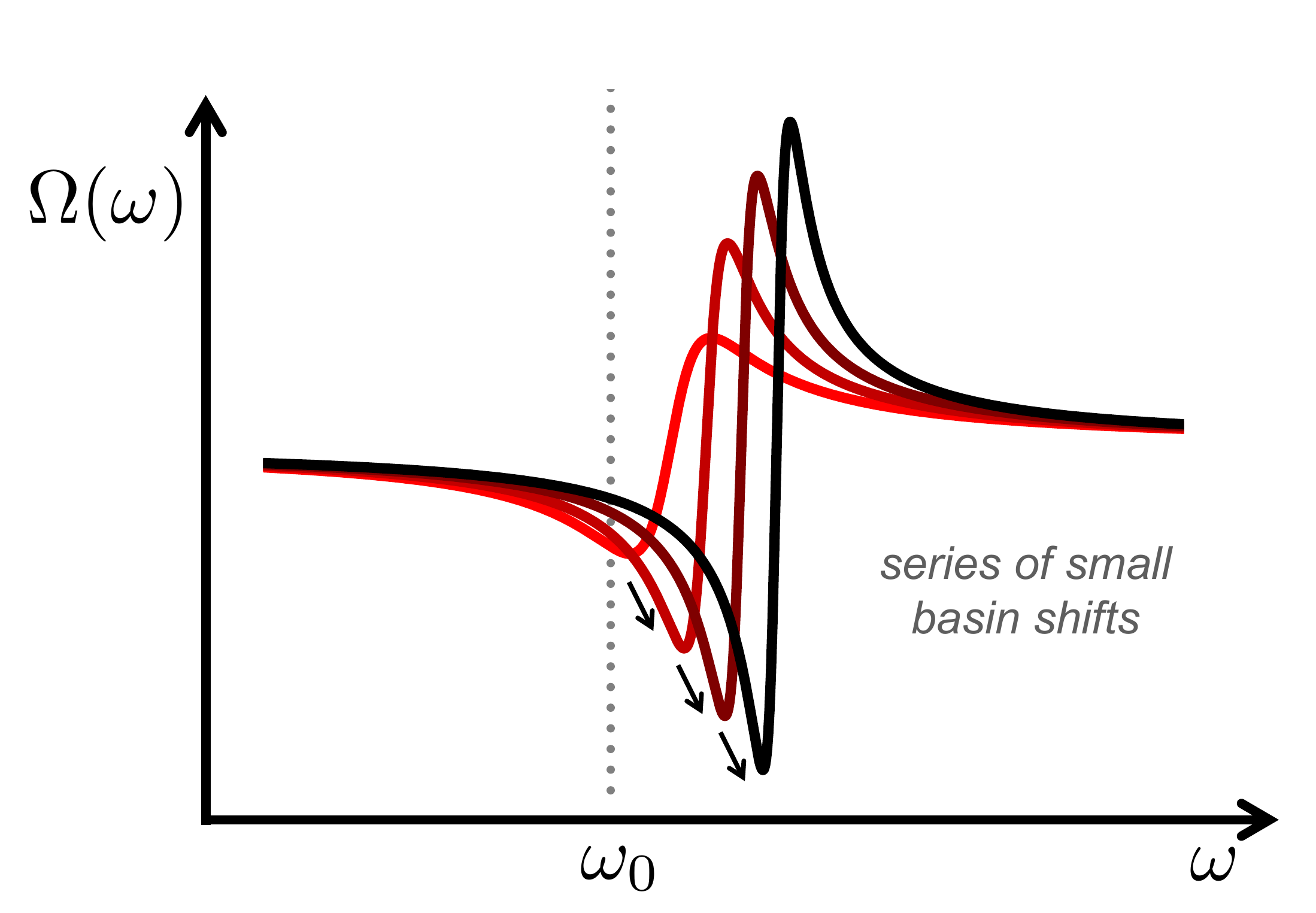}
\caption{Example of a series of small basin shifts that occur during the transitional-$\omega$ stage of the optimization.
        }
\label{fig:small_basin_shifts}
\end{figure}

In practice, one must take care in handling the transition between the initial ``fixed-$\omega$'' and the final ``adaptive-$\omega$'' stages
of the optimization.
When $\omega$ is changed, the nonlinear function being optimized is altered.
One can imagine that a large alteration made abruptly at the end of the fixed-$\omega$ stage (Figure \ref{fig:large_basin_shift})
could move the function's minimum far enough that the wave function variables were no longer within the basin of convergence for
the chosen update method.
Indeed, we have observed exactly this behavior in some tests involving $\Omega$ and the linear method update scheme.
To avoid such pathologies, we instead add a ``transitional-$\omega$'' stage to our optimization, in which $\omega$ is gradually
interpolated between its initial fixed value and the value required to achieve variance minimization.
In this way, the variational principle's minimum is moved only small steps at a time (Figure \ref{fig:small_basin_shifts})
to ensure that the wave function remains within its basin of convergence.
In this study, we use the interpolation
\begin{align}
\label{eqn:shift_update_period}
\omega_j &= \alpha_j \hspace{0.6mm} \omega_0 + ( 1-\alpha_j ) (E_{j-1} - \sigma_{j-1}) \\
\label{eqn:shift_update_period_alpha_k}
\alpha_j &= \begin{cases} 1                         \hspace{3.80cm} j \le N_F \\
                          \frac{1}{N_T}(N_F+N_T-j)  \hspace{0.70cm} N_F < j \le N_F+N_T \\
                          0                         \hspace{3.78cm} j > N_F+N_T
              \end{cases}
\end{align}
in which $j$ is the linear method iteration number and $N_F$ and $N_T$ are the number of iterations in the fixed-$\omega$ and
transitional-$\omega$ stages, respectively.
In the cases tested here, we find that $N_F$ between 5 and 20 and $N_T$ between 10 and 20 are effective choices.

In addition to ensuring both state selectivity and size consistency, this strategy improves the
practical usability of excited state variational principles in VMC.
Although the final wave function's energy is often not very sensitive to the choice of $\omega$ \cite{Zhao:2016:dir_tar},
there are likely to be cases where the user's choice of $\omega$ has a meaningful effect on the results.
In previous work, $\omega$ has sometimes been adjusted by hand in order to minimize $\Omega$.
Although this does make the choice of $\omega$ unique, the process is tedious and prevents the overall methodology from
achieving black box operation.
With the adaptive approach described here, a user need only specify the initial value $\omega_0$ so as to target the
desired state.

\section{Results}
\label{sec:results}

\subsection{Computational Details}

In the next few sections, we will present numerical results that complement our formal analysis.
Results for CO and N$_2$ were obtained with our own Hilbert space VMC software in a STO-3G basis \cite{Pople:1969:sto-3g},
with integrals imported from PySCF \cite{Pyscf_brief}.
Bond distances were fixed at 1.19 {\AA} and 1.18 {\AA} for CO and N$_2$, respectively.
For the formaldehyde-water system, the geometry was optimized to a local minimum (see Figure \ref{fig:formaldehyde_water_structure})
using the $\omega$-B97X-D density functional \cite{chai:2008:wb97xd} and a 6-311G basis \cite{Pople:1980:6311g} set within QChem \cite{shao:2015:qchem}.
VMC results for this system were obtained with a development version of \uppercase{QMCPACK} \cite{Kim:2012:qmcpack_scaling} with molecular orbitals and
configuration interaction singles (CIS) \cite{dreuw2005single} initial guesses imported from \uppercase{GAMESS} \cite{gamess}.
Equation of motion coupled cluster with singles and doubles (EOM-CCSD) \cite{Krylov:2008:eom_ccsd} results were obtained with \uppercase{MOLPRO} \cite{MOLPRO_web,MOLPRO_paper}.
The VMC orbitals as well as the CIS and EOM-CCSD results for the formaldehyde-water system used the pseudopotentials of Burkatzki et al. \cite{Burkatzki:2007:pseudopot}, replacing core electrons for C and O atoms, and the corresponding valence double zeta (VDZ) basis set. 

\begin{figure}[h]
\includegraphics[width=3.33in]{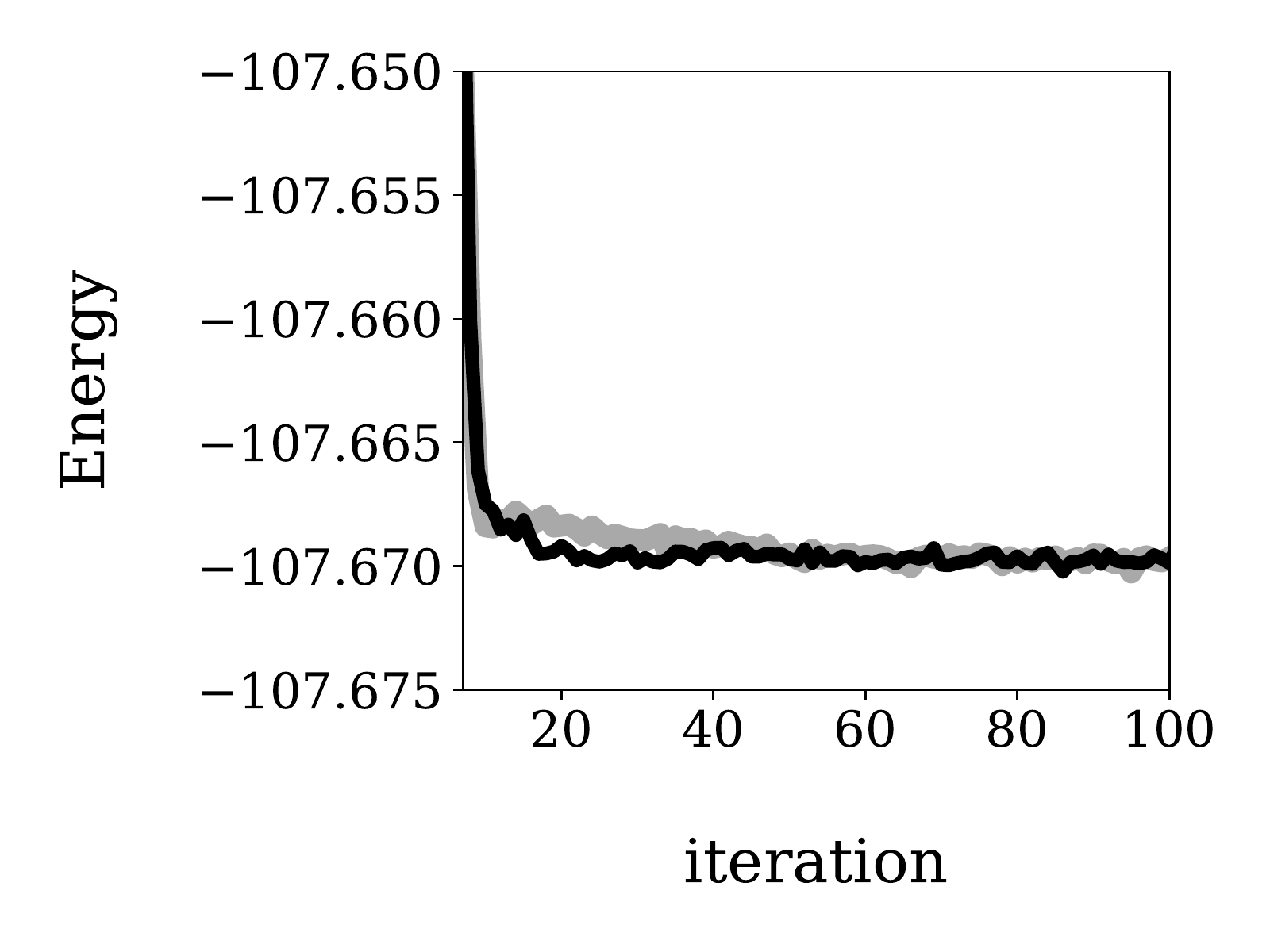}
\caption{Energy in a.u.\ with respect to optimization step for the ground state of N$_2$ optimized with $\sigma^2$ minimization (gray) and our adaptive-$\omega$ method (black).}
\label{fig:tf_vm_N2_Energy}
\end{figure}

\begin{figure}[h]
\includegraphics[width=3.33in]{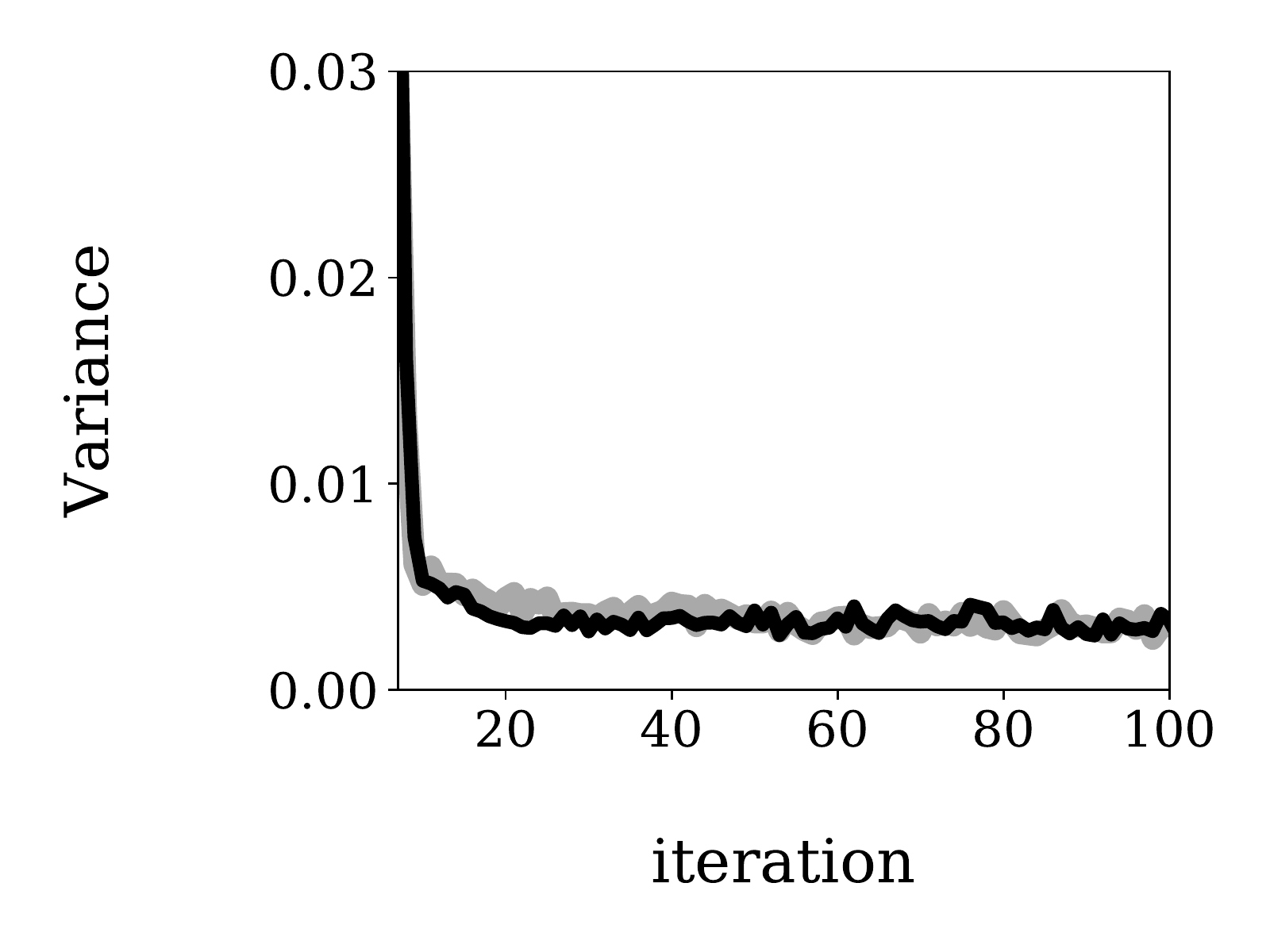}
\caption{Variance in a.u.\ with respect to optimization step for the ground state of N$_2$ optimized with $\sigma^2$ minimization (gray) and our adaptive-$\omega$ method (black).}
\label{fig:tf_vm_N2_Variance}
\end{figure}

\subsection{Comparison to Variance Minimization}
\label{sec:compare_to_variance}

\begin{figure}[b]
\includegraphics[width=3.33in]{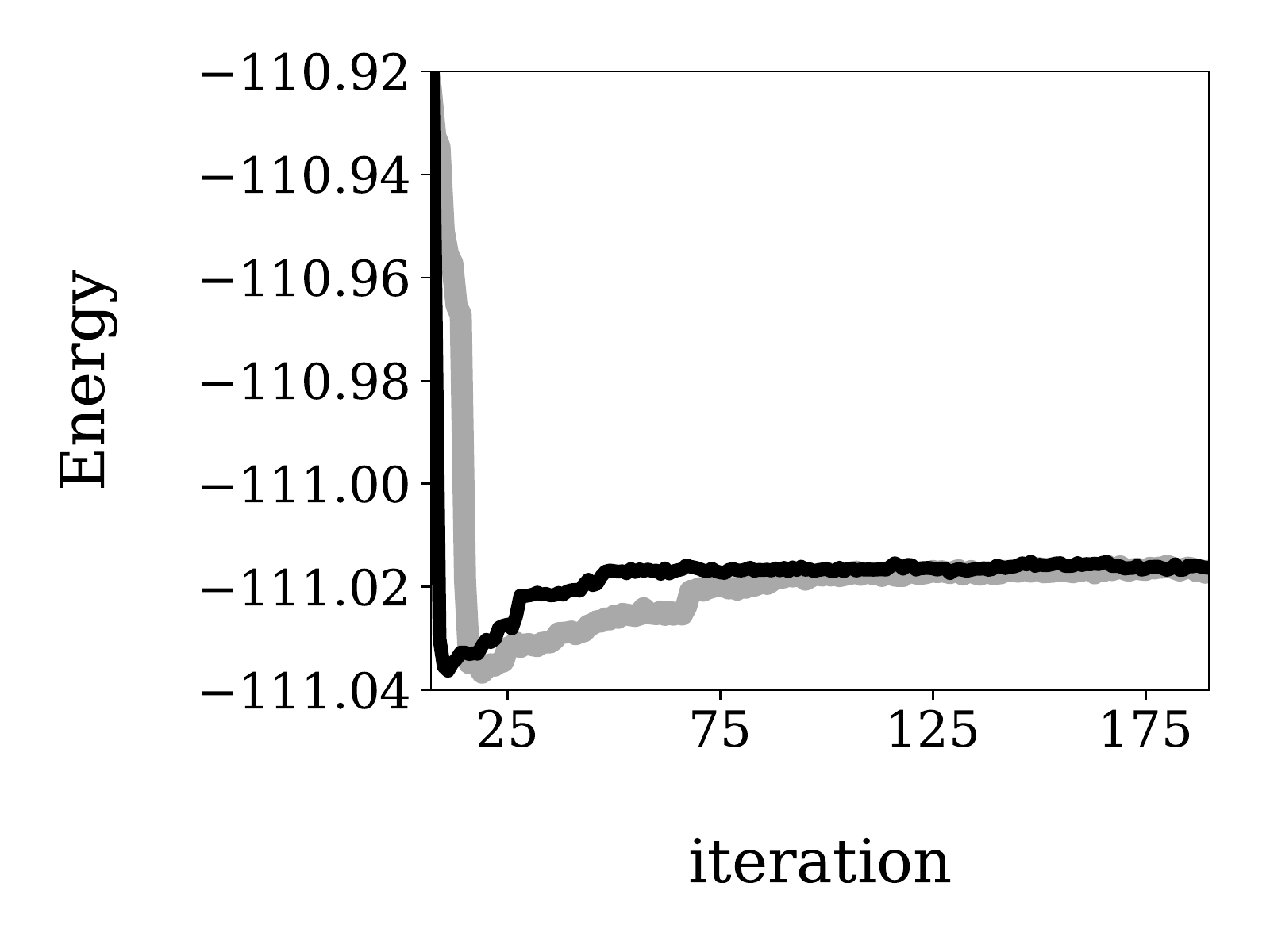}
\caption{Energy in a.u.\ with respect to optimization step for the first excited singlet of CO optimized with $\sigma^2$ minimization (gray) and our adaptive-$\omega$ method (black).}
\label{fig:tf_vm_CO_Energy}
\end{figure}

\begin{figure}
\includegraphics[width=3.33in]{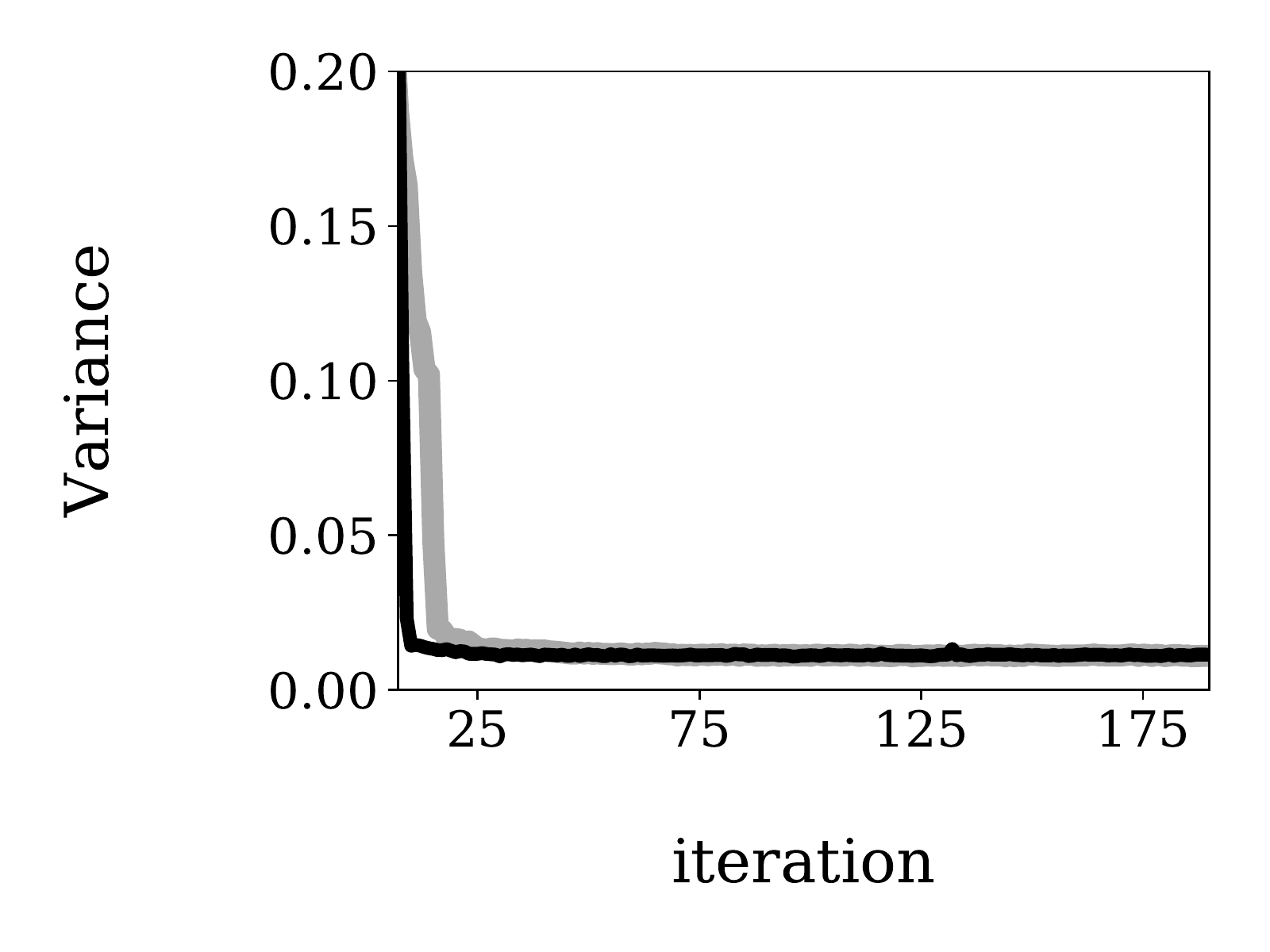}
\caption{Variance in a.u.\ with respect to optimization step for the first excited singlet of CO optimized with $\sigma^2$ minimization (gray) and our adaptive-$\omega$ method (black).}
\label{fig:tf_vm_CO_Variance}
\end{figure}

In this section, we use the JAGP ansatz in Hilbert space to compare the results of our optimization scheme to those that are obtained by a simple
minimization of $\sigma^2$.
To begin, we apply both optimizations (with $\omega_0=-109.00$ Hartrees, $N_F=8$, and $N_T=10$ for the adaptive-$\omega$ method) to the ground state of N$_2$,
using the restricted Hartree-Fock (RHF) determinant with slightly randomized orbital coefficients for the JAGP initial guess.
As seen in Figures \ref{fig:tf_vm_N2_Energy} and \ref{fig:tf_vm_N2_Variance}, the two methods produce the same values for $E$ and $\sigma^2$
upon convergence.
We also apply both optimization methods (with $\omega_0=-111.30$ Hartrees, $N_F=15$, and $N_T=10$ for the adaptive-$\omega$ method) to the first excited singlet
of CO.
In this case, the initial guess for the JAGP pairing matrix was constructed by adding a HOMO-LUMO promotion and slight orbital coefficient randomization
to the RHF ground state to produce a crude open-shell singlet representation.
Figures \ref{fig:tf_vm_CO_Energy} and \ref{fig:tf_vm_CO_Variance} show that, starting from this guess, both simple variance minimization and
our $\Omega$-based, adaptive-$\omega$ method converge to the same result, as expected.

\begin{figure}
\includegraphics[width=3.33in]{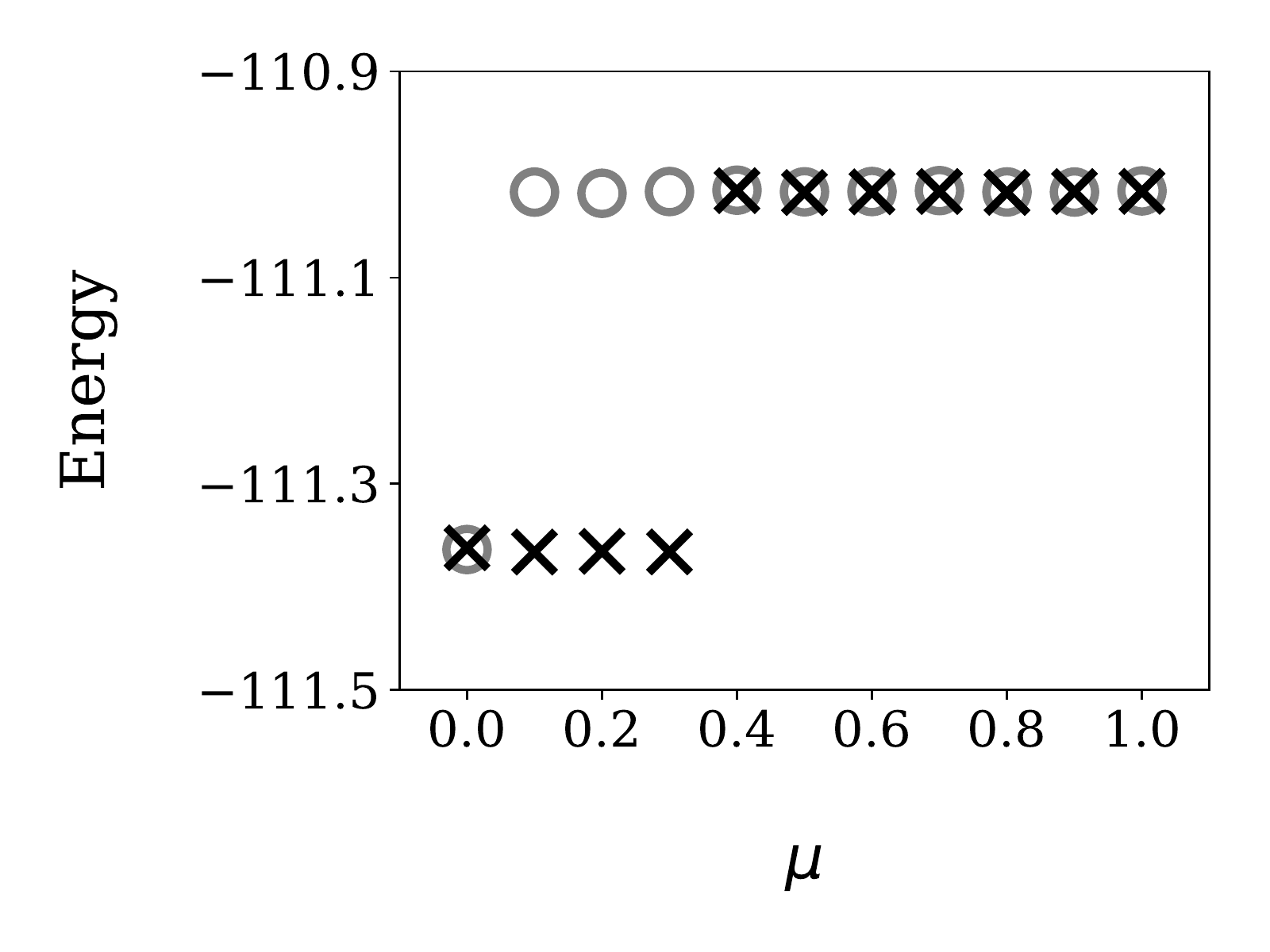}
\caption{Energies of optimized JAGP wave functions for CO when starting from a range of different initial guesses that interpolate between a pure ground state guess ($\mu=0$)
         and a pure excited state guess ($\mu=1$).
         Results are shown for both $\sigma^2$ minimization (black X) and our adaptive-$\omega$ method (gray circles).}
\label{fig:direct_targeting}
\end{figure}

To compare the state selectivity of our method with that of variance minimization, we have also performed a series of optimizations for CO in which the initial guess
for the wave function was interpolated between a ground state guess and an excited state guess.
Specifically, we used pairing matrix guesses of the form
\begin{align}
\label{eqn:qmc_omega_schemes}
\mathcal{M} = ( 1 - \mu ) \mathcal{M}^{(0)} + \mu \hspace{0.7mm} \mathcal{M}^{(1)}
\end{align}
where $\mathcal{M}^{(0)}$ is the pairing matrix corresponding to the RHF ground state,
$\mathcal{M}^{(1)}$ is the open-shell singlet pairing matrix resulting from a HOMO-LUMO promotion,
and $\mu \in [0,1]$.
As shown in Figure \ref{fig:direct_targeting}, simple variance minimization converged to the ground state when $\mu \leq 0.3$.
In contrast, our $\Omega$-based optimization (with $\omega_0 = -111.30$ Hartrees, $N_F=20$, and $N_T=10$) converged to the (targeted) excited state for all cases except $\mu=0$.
Indeed, we found that as little as 0.5\% excited state character (i.e.\ $\mu=0.005$) in the initial guess was sufficient for our method to converge to the excited state,
providing a clear example of the advantage offered by a state selective approach.

\subsection{Size Consistency}

As shown in Section \ref{sec::proof_no_sc}, optimizing $\Omega$ with a particular, fixed value of $\omega$ can lead to size consistency
issues when working with approximate wave functions.
Of course, with exact wave functions, results will be $\omega$-independent and size consistency will be achieved because in this limit,
minimizing $\Omega$ will produce exact Hamiltonian eigenstates \cite{Zhao:2016:dir_tar}. 
While it is not always easy to tell how far from this limit one is, one indication may be how sensitive the optimized wave function's
energy is to the precise choice of $\omega$.
In the systems at hand, we find that N$_2$'s ground state energy is quite insensitive to $\omega$, varying by less than $10^{-3}$ Hartrees for
fixed-$\omega$ optimizations in which $\omega$ is set anywhere between $-107.7$ to $-108.6$ Hartrees.
The first excited state of CO is more sensitive, with fixed-$\omega$ optimizations producing energies that change by as much as 0.01 Hartrees as
$\omega$ is varied between $-111.10$ and $-111.35$ Hartrees.

Given that the JAGP wave function approximation does produce nontrivial sensitivity to $\omega$ in at least one of these molecules,
it is an interesting case in which to investigate size consistency.
We should stress that, although it is an approximate ansatz, the Hilbert-space JAGP product factorizes and so will produce size consistent
energies when paired with a size consistent variational principle \cite{Neuscamman:2012:sc_jagp}.
Thus, any size consistency violation in its use can be linked to the variational principle.
The test we perform is to optimize N$_2$'s ground state and CO's first excited state, both separately and when the two molecules
are treated together at a distance of 20 {\AA}.
This test provides a simple case in which we may ask whether CO's excited state is affected by the presence of a far away molecule.

When we minimize $\Omega$ with a fixed value of $\omega$ chosen in Hartrees as $\omega=\tilde{E}- 0.26$, with $\tilde{E}$
being the expected energy based on variance minimization results for the separate molecules,
we find that the size consistency error, $E_{CO+N_2}-E_{CO}-E_{N_2}$, is over 2 milliHartrees.
When instead we employ our adaptive-$\omega$ method, the results are size consistent to within our statistical uncertainty,
as shown in Table \ref{tab:size_con_err}.
Thus, the ability to gradually transform the variational principle so that
it is equivalent to variance minimization at convergence allows size consistent results to be achieved.

\begin{table}[h]
\centering
\caption{Size consistency errors $|E_{CO+N_2}-E_{CO}-E_{N_2}|$ and their statistical uncertainties for the first singlet excited state of CO when combined with a far away nitrogen molecule.
        }
\label{tab:size_con_err}
\begin{tabular}{ c  l@{ }l@{ }l }
\hline\hline
\hspace{0mm} Method \hspace{0mm} & 
\multicolumn{3}{ c }{ \hspace{0mm} Error (mE$_h$) \hspace{0mm} } \\
\hline
 \hspace{0mm} fixed-$\omega$    \hspace{0mm} & \hspace{0mm} 2.6  \hspace{0.0mm} & $\pm$ \hspace{0.0mm} & 0.2 \\
 \hspace{0mm} adaptive-$\omega$ \hspace{0mm} & \hspace{0mm} 0.04 \hspace{0.0mm} & $\pm$ \hspace{0.0mm} & 0.2 \\
\hline\hline
\end{tabular}
\end{table}

\subsection{Formaldehyde and water}

\begin{figure}
\includegraphics[width=3.33in]{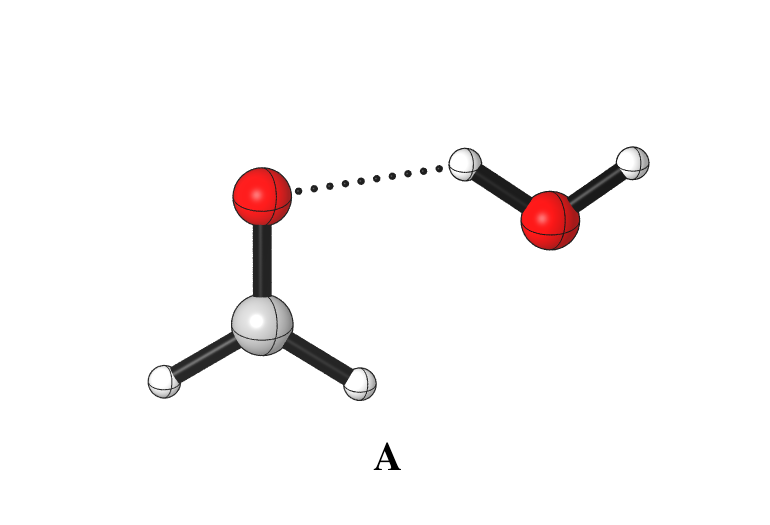}
\caption{Hydrogen-bonded formaldehyde-water complex. \cite{CYLView} }
\label{fig:formaldehyde_water_structure}
\end{figure}

To show a slightly more realistic example where having a size consistent optimization method matters, we turn
to a hydrogen-bonded complex between formaldehyde and water, shown as complex \textbf{A} in Figure \ref{fig:formaldehyde_water_structure}.
To evaluate the first singlet excitation energy on the formaldehyde, we employ the recently-developed
variation after response (VAR) approach \cite{Neuscamman:2016:var,Blunt:2017:fdlr} as implemented in a development version
of QMCPACK for a Slater determinant in real-space.
In this context, VAR uses a finite-difference scheme to allow orbital optimization and a Jastrow factor to be applied variationally to
a CIS-like linear response expansion \cite{Blunt:2017:fdlr}.
By combining the adaptive-$\omega$ optimization we've presented here with VAR's ability to start from
the output of a CIS calculation, we intend to show how VMC may deal with an excited state in a relatively black box fashion.

\begin{figure}
\includegraphics[width=3.33in]{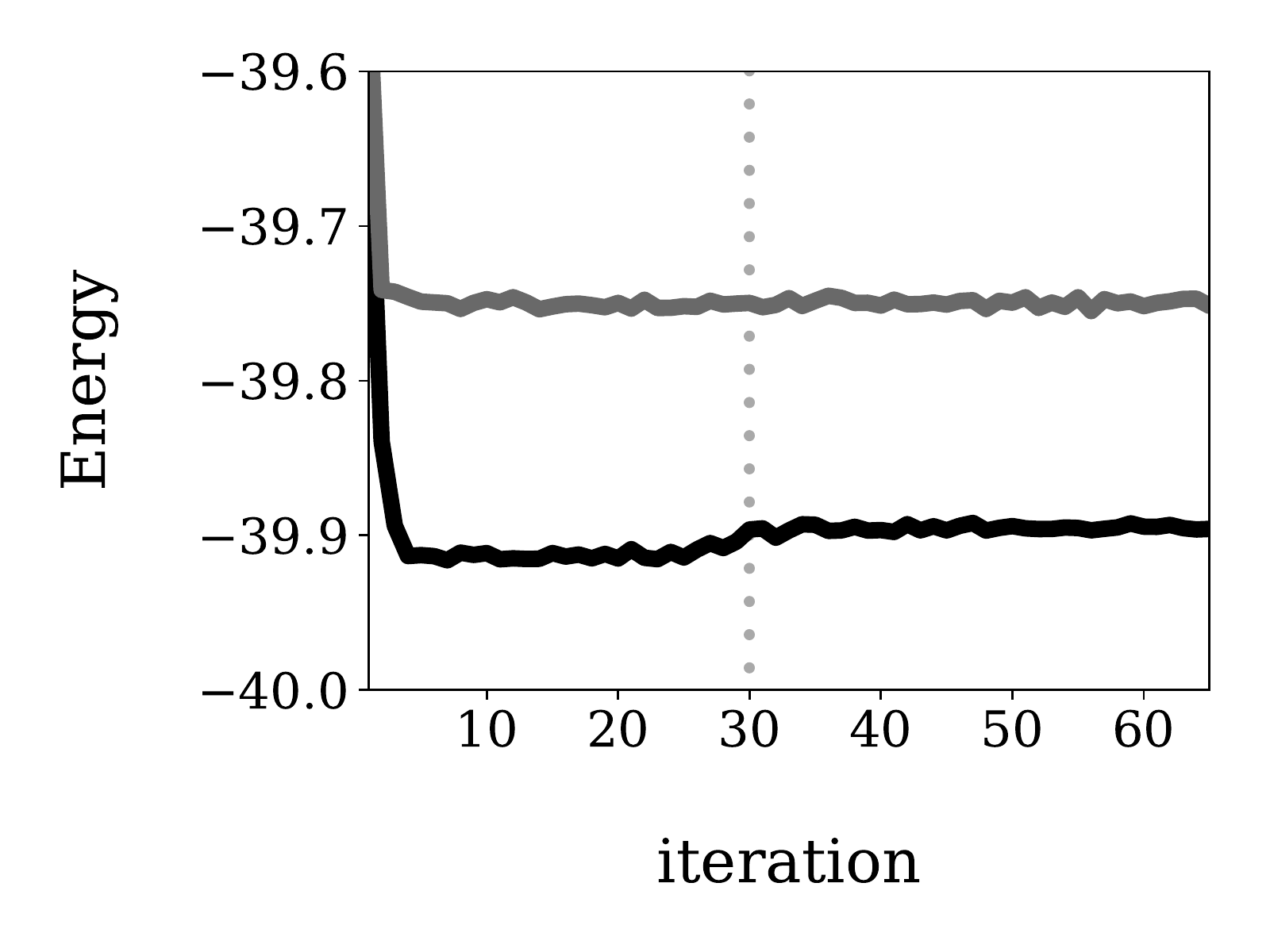}
\caption{Energy versus iteration for the ground state (solid black) and excited state (solid gray) of \textbf{A}.
         The dotted line marks the end of the transitional-$\omega$ phase.
        }
\label{fig:formaldehyde_optimization_E}
\end{figure}

For the Jastrow-Slater ground state optimization, we chose $\omega_0 = -44.175$ to be several Hartrees below the RHF energy.
As shown in Figure \ref{fig:formaldehyde_optimization_E}, this resulted in the energy increasing slightly during the transitional-$\omega$ phase
of the optimization, as is to be expected when converting from a ground-state-energy-like variational principle to something more akin to variance minimization.
For the VAR excited state optimization, we chose $\omega_0=-41.175$ Hartrees, which is in between the values $-40.8747$ and $-42.2812$ that result for $E_{CIS}-\sigma$
when $\sigma$ is taken from the optimized Jastrow-Slater ground state or the CIS wave function, respectively.
For both states, we set $N_F=10$ and $N_T=20$.

\begin{table}[h]
\centering
\caption{Excitation energies in E$_h$ for complex \textbf{A}.}
\label{tab:formaldehyde_qmc}
\begin{tabular}{ c  r@{.}l@{ }l@{ }l  }
\hline\hline
\hspace{0mm} Method \hspace{0mm}  & 
\multicolumn{4}{ c }{ \hspace{0mm} Excitation Energy \hspace{0mm} } \\
\hline
 \hspace{0mm} CIS       \hspace{0mm} & \hspace{6mm}  0&1718   &       &     \\
 \hspace{0mm} EOM-CCSD  \hspace{0mm} & \hspace{6mm}  0&1511   &       &     \\
 \hspace{0mm} VMC       \hspace{0mm} & \hspace{6mm}  0&1439   & $\pm$ & 0.0006 \\
\hline\hline
\end{tabular}
\end{table}

As shown in Table \ref{tab:formaldehyde_qmc}, the VMC-based excitation energy agrees more closely with that of EOM-CCSD than
with that of CIS.
Presumably, this is due to the excited state orbital optimization lowering the excited state energy as compared to CIS,
whereas even in RHF the ground state already enjoyed state specific orbital optimization and so its energy was lowered less by VMC.
Although complex \textbf{A} is used here for the purposes of demonstration, the ability of VMC to produce a relatively accurate, nearly black box
result for an excited state in the presence of a hydrogen-bonded solvent molecule is promising.
Given QMC's low scaling compared to EOM-CCSD ($N^4$ versus $N^6$), it will be interesting to explore its prospects in larger and more
technologically relevant examples of solvated photo-absorbers.

\section{Conclusions}
\label{sec:conclusions}

We have shown that size consistency is lacking in interior state selective variational principles
that are analytic around their global minima and based on at most the second power of the Hamiltonian,
a set we have denoted as $V_{1,2}$.
In contrast, the well-established approach of variance minimization is known to be size consistent but
not state selective.
To achieve the best of both worlds, we have proposed a general optimization strategy that amalgamates
variance minimization with a state selective variational principle from $V_{1,2}$.
The approach is size consistent at convergence and maintains rigorous state selectivity at all stages.
In an initial exploration with the $\Omega$ variational principle, we find that it is important that
the transformation of the variational principle proceed gradually, lest the basin of convergence
be moved away from the current wave function.
We note that the overall strategy is applicable to a wide range of variational principles,
including the $W$ function employed recently by $\sigma$-SCF, and is readily
compatible with the leading wave function optimizers in variational Monte Carlo.

Having demonstrated a viable path to variational, size consistent excited states, it is worth
considering where such methods may be most useful in future.
Thanks to QMC's ability to work with either open or periodic boundary conditions, the methodology
should be equally applicable in both solids and molecules.
Thus, in addition to its implications for modeling charge transfer excitations in the presence
of explicit solvent, the approach may also prove useful in modeling defect-centered excitons.
Regardless of the specific application, the removal of 
$\omega$ as a free parameter should make QMC-based excited state
investigations more straightforward.
In conjunction with recent variation-after-response developments that can build atop simple excited
state quantum chemistry methods, the optimization approach presented here presages a more
black box route to high-accuracy QMC results in a wide variety of excited state applications.

\section{Acknowledgments}

This work was supported by the U.S. Department of Energy, Office of Science, Basic Energy Sciences,
Materials Sciences and Engineering Division, as part of the Computational Materials Sciences Program
and Center for Predictive Simulation of Functional Materials.
Calculations were performed using the Berkeley Research Computing Savio cluster. 


\section{Appendix A}

Here we show that for any state specific target function $\Gamma(E,\sigma^2)$ that is real analytic
(i.e.\ can be written as a convergent power series with real coefficients)
in a region around its global minimum, there exist system/ansatz pairs for which neither
$E$ nor $\sigma^2$ is stationary at the $\Gamma$ minimum.
We will do so by constructing a particular counterexample, although we suspect that other counterexamples exist.
Consider a system in which three of the Hamiltonian eigenvalues are $-b$, $1$, and $2$, with normalized eigenvectors
\begin{align}
\hat{H}|\Phi_b\rangle  = -b|\Phi_b\rangle, \quad
\hat{H}|\Phi_1\rangle  =   |\Phi_1\rangle, \quad \mathrm{and} \quad
\hat{H}|\Phi_2\rangle  =  2|\Phi_2\rangle.
\label{eqn:three_level_states}
\end{align}
For our approximate ansatz we choose the single-variable wave function 
\begin{align}
|\Psi(x)\rangle = x |\Phi_b\rangle + |\Phi_1\rangle + (b^2+x)|\Phi_2\rangle,
\label{eqn:three_level_wfn}
\end{align}
in which $x$ is allowed to take on real values.
This defines a set of system/ansatz pairs in which we can control through $b$ how closely the ansatz can come to
an exact representation of the $|\Phi_1\rangle$ eigenstate, which we will take to be the state targeted
by $\Gamma$.

The energy and variance of this ansatz can be written as
\begin{align}
\label{eqn:three_level_e}
E & = \frac{P}{D},  \\
\label{eqn:three_level_v}
\sigma^2 &= \frac{QD - P^2}{D^2},
\end{align}
in which we have used the three polynomials
\begin{align}
\label{eqn:three_poly_1}
P & = 1 - b x^2 + 2(b^2+x)^2, \\
\label{eqn:three_poly_2}
Q & = 1 + b^2 x^2 + 4(b^2+x)^2, \\
\label{eqn:three_poly_3}
D &= 1 + x^2 + (b^2+x)^2.
\end{align}
Using the properties of the geometric series and the fact that $D$ cannot be zero when $b>0$,
we note that both $E$ and $\sigma^2$ are analytic functions of $x$ and $b$ so long as
the point $(x,b)$ is sufficiently close to $(0,0)$.
By inspecting the stationary points of $E$ and $\sigma^2$, we will find that we can always choose
$b$ positive but small enough that these stationary points are distinct from the $\Gamma$ minimum.

Begin with the stationary point for the energy, at which
\begin{align}
\label{eqn:three_en_stat}
\frac{\partial E}{\partial x}=0,
\end{align}
which may be rearranged as
\begin{align}
\label{eqn:three_en_stat_rearr}
b \Big( \hspace{0.5mm} (2b+b^2)x^2 + (1+2b^3+b^4)x - b \hspace{0.5mm} \Big)= 0,
\end{align}
from which we see that the energy is always stationary when $b=0$.
When $b$ is small but positive, we will have two roots, but only one of them,
\begin{align}
\label{eqn:x_e}
x_E = b - 2b^3 + \mathcal{O}(b^4),
\end{align}
will occur near the origin.
The other root,
\begin{align}
\label{eqn:x_b}
x_F = - \frac{1}{2b} + \frac{1}{4} + \mathcal{O}(b),
\end{align}
will be far from the origin.

Moving on to the variance, we find that its stationary condition,
\begin{align}
\label{eqn:three_var_stat}
\frac{\partial \sigma^2}{\partial x}=0
\end{align}
can be rearranged into a cubic polynomial in $x$,
\begin{align}
0 =&   2b^2(1-b^4)
     + 2\big(2+b(2+b+6b^4+2b^5+4b^7+4b^8+b^9)\big) x \notag \\
   & + 6b^2 (1+b)^3 \big(2+b(b^2+b-2)\big)x^2
     + 4\big(2+2b+b^4(2+b)^2\big)x^3.
\label{eqn:three_var_stat_rearr}
\end{align}
At small values of b, one can show that the discriminant of this
polynomial is negative, implying that it has one real and two
complex roots.
As our ansatz does not admit complex values for $x$, the variance
will have only one stationary point.
Using the cubic formula and assuming $b$ is small, this root
can be found to be
\begin{align}
\label{eqn:x_v}
x_V = -\frac{b^2}{2} + \frac{b^3}{2} - \frac{b^4}{4} + \mathcal{O}(b^6).
\end{align}

Having found the energy and variance stationary points,
we now consider the target function as given in Eq.\ (\ref{eqn:gamma_taylor_3}).
For our particular system/ansatz choice, we can see that $\Gamma$ will
be a real analytic function of $x$ and $b$ when $b$ is small and $x$ is close to
the global minimum, which we know occurs at $x=0$ when $b$ is chosen to be $0$.
This is because both $E$ and $\sigma^2$ are analytic in $x$ and $b$ in this
region, and by assumption $\Gamma$ is real analytic near its global minimum.
Furthermore, for small $b$ and $x$, we have
\begin{align}
\label{eqn:ev_at_b_eq_0}
\Delta   &= b \hspace{0.7mm} ( \hspace{0.7mm} b^3 + 2 b x - x^2 + \mathrm{higher~order~terms} \hspace{0.7mm} ), \\
\sigma^2 &= b^4 + 2 b^2 x + (2+2b+b^2) x^2 + \mathrm{higher~order~terms},
\end{align}
and so at $b=0$, the leading order term in $\Gamma$ will be an
even power of $x$ with degree $2$ or higher.
This makes sense as the minimum could not be $x=0$ if the leading term were odd.
Let us now express the stationary condition for $\Gamma$ as
\begin{align}
\label{eqn:gamma_stat_cond}
\frac{\partial\Gamma}{\partial x} = Z(x,b) = 0.
\end{align}
By the differentiability of analytic functions, $Z$ will be real analytic and thus smooth
in the region surrounding $\Gamma$'s minimum.
When $b=0$, we know that $Z$ will have an odd power of $x$ with degree $1$ or greater
as its leading order term,
and so the stationary point that is the global minimum will occur at $x=0$ as expected.
As $Z$ is a smooth function of both $x$ and $b$, this implies that when $b>0$, the
value of $x$ that minimizes $\Gamma$ can be forced arbitrarily close to zero by
making $b$ sufficiently small.
Finally, note that $Z(x,0)$ is a nonconstant function of $x$ in the region of small $x$,
as this is required for the global minimum of $\Gamma$ to be unique.
As $Z$ is smooth in $b$, this implies that if we hold $b$ fixed at a positive but small value,
$Z$ will still be a nonconstant function of $x$.
By the principle of permanence, this function will have isolated roots, and so there
will be only one stationary point of $\Gamma$ that approaches $x=0$ as $b$ becomes small.
It now remains to show that $b$ can be chosen positive but small enough so as to prevent this
stationary point coinciding with either $x_E$ or $x_V$, which are the only stationary
points of the energy and variance that approach $0$ as $b$ becomes small.

First consider $x_E$.
When $b$ is small and $x=x_E$, we find that
\begin{align}
\label{eqn:ev_at_xe}
\Delta = b^3 + b^4 + \mathcal{O}(b^5) \quad \mathrm{and} \quad \sigma^2 = 2b^2 + 4b^3 + \mathcal{O}(b^4).
\end{align}
This implies that for $x=x_E$ and small but nonzero $b$, $\sigma^2$ will be small and nonzero.
Furthermore, because $x_E$ and $x_V$ are distinct for nonzero $b$, $\sigma^2$ will not be stationary here.
We will now deal with two cases for $\Gamma$ separately.
First, if the middle sum in Eq.\ (\ref{eqn:gamma_taylor_3}) is absent, we have
\begin{align}
\label{eqn:qr_inf_at_xe}
\frac{\partial\Gamma}{\partial x}
= \sum_{m=p}^\infty m \hspace{0.5mm} a_{m0} \hspace{0.5mm} (\sigma^2)^{m-1} \frac{\partial\sigma^2}{\partial x},
\end{align}
which, based on what we know about $\sigma^2$, shows that $\Gamma$ will not be stationary at $x=x_E$
if $b$ is positive and sufficiently small.
For the second case, in which the middle sum is present, we have to be more careful.
Using the fact that $\sigma^2$ is not stationary at $x=x_E$, the stationary condition
$\partial\Gamma/\partial x=0$ can for this second case be written as
\begin{align}
\label{eqn:qr_fin_at_xe}
0 = \sum_{m=p}^\infty m \hspace{0.5mm} a_{m0} \hspace{0.5mm} (\sigma^2)^{m-1}
  + \sum_{m=q}^\infty \sum_{n=r}^\infty m \hspace{0.5mm} a_{mn} \hspace{0.5mm} (\sigma^2)^{m-1} \Delta^n.
\end{align}
If $p\le q$, this condition cannot be satisfied at small but nonzero $b$ as all the terms in the right hand sum
will be higher order in $b$ than the first term in the left hand sum, because $r>0$ and $\Delta$ is order $b^3$.
If instead $p>q$, we may divide through by $(\sigma^2)^{q-1}$ to obtain
\begin{align}
\label{eqn:qr_fin_at_xe_2}
0 = \sum_{m=p}^\infty m \hspace{0.5mm} a_{m0} \hspace{0.5mm} (\sigma^2)^{m-q}
  + \sum_{m=0}^\infty \sum_{n=r}^\infty (m+q) a_{(m+q)n} \hspace{0.5mm} (\sigma^2)^{m} \Delta^n.
\end{align}
This equation is analytic in $b$ and has a solution at $b=0$.
As nonconstant analytic functions have isolated roots, 
Eq.\ (\ref{eqn:qr_fin_at_xe_2}) cannot also have a solution at arbitrarily small but positive
$b$ unless the right hand side is zero for all $b$.
We can show that this is not so by inspecting the leading order terms in $b$.
Using Eq.\ (\ref{eqn:ev_at_xe}), we see that if the lowest order terms from the two sums are to be the same order,
we must have
\begin{align}
\label{eqn:pqr_condition}
2(p-q) = 3r,
\end{align}
which implies that there is a positive integer $z$ such that
\begin{align}
\label{eqn:pqrz_condition}
p-q = 3 z \quad \mathrm{and} \quad r = 2z.
\end{align}
If so, the leading order terms from Eq.\ (\ref{eqn:qr_fin_at_xe_2})'s left hand sum will be proportional to
\begin{align}
\label{eqn:xe_leading_lhs}
b^{6z}+6zb^{6z+1}
\end{align}
while the leading order terms from its right hand sum will be proportional to
\begin{align}
\label{eqn:xe_leading_lhs}
b^{6z}+2zb^{6z+1}.
\end{align}
Thus, regardless of the values of $p$, $q$, and $r$, these two sums cannot cancel exactly
and so $\Gamma$ will not be stationary at $x=x_E$ when $b$ is chosen to be positive and small.

We follow a similar analysis to show that $\Gamma$ will not be stationary at $x=x_V$, where
\begin{align}
\label{eqn:ev_at_xv}
\Delta = \frac{3b^5}{4} - \frac{b^7}{2} + \mathcal{O}(b^8)
\quad \mathrm{and} \quad
\sigma^2 = \frac{b^4}{2} + \frac{b^5}{2} + \mathcal{O}(b^6).
\end{align}
As the energy is not stationary at $x=x_V$, the stationary condition $\partial\Gamma/\partial x=0$ can
be written as
\begin{align}
\label{eqn:qr_fin_at_xv}
0 = \sum_{m=q}^\infty \sum_{n=r}^\infty n \hspace{0.5mm} a_{mn} \hspace{0.5mm} (\sigma^2)^{m} \Delta^{n-1}
  + \sum_{n=s}^\infty n \hspace{0.5mm} a_{0n} \hspace{0.5mm} \Delta^{n-1}.
\end{align}
If the $qr$ sum is not present, then clearly $\Gamma$ will not be stationary at $x=x_V$ when $b$ is positive and small.
Otherwise, following the same logic we used for $x_E$, we now need to show that the right hand side of
Eq.\ (\ref{eqn:qr_fin_at_xv}) is not zero for all $b$.
If $s\le r$, the leading order term in the right hand sum will be of a different order in $b$ than that of the
left hand sum, and so the right hand side will be nonzero for $b$ small and positive.
If instead $s>r$, we may divide through by $\Delta^{r-1}$ to obtain
\begin{align}
\label{eqn:qr_fin_at_xv_2}
0 = \sum_{m=q}^\infty \sum_{n=0}^\infty (n+r) \hspace{0.5mm} a_{m(n+r)} \hspace{0.5mm} (\sigma^2)^{m} \Delta^{n}
  + \sum_{n=s}^\infty n \hspace{0.5mm} a_{0n} \hspace{0.5mm} \Delta^{n-r}.
\end{align}
If the two sums are to cancel, so must their leading order terms.
Using Eq.\ (\ref{eqn:ev_at_xv}), we see that this can only occur if
\begin{align}
\label{eqn:qrs_condition}
4q = 5(s-r),
\end{align}
which implies that there is a positive integer $y$ such that
\begin{align}
\label{eqn:qrsz_condition}
q = 5 y \quad \mathrm{and} \quad s-r = 4y.
\end{align}
If $q$, $r$, and $s$ have this relationship, we find that the left hand sum is proportional to
\begin{align}
\label{eqn:xv_left_sum}
b^{20y} + 5yb^{20y+1} + \mathcal{O}(b^{20y+2})
\end{align}
while the right hand sum is proportional to
\begin{align}
\label{eqn:xv_right_sum}
b^{20y} - \frac{8y}{3} b^{20y+2} + \mathcal{O}(b^{20y+3}).
\end{align}
Thus, regardless of the values of $q$, $r$, and $s$, we see that $\Gamma$ will not be stationary
at $x=x_V$ when $b$ is positive and small.

To conclude, we find that in this system/ansatz pairing, $b$ can be chosen to be positive but small enough
such that the minimum of $\Gamma$, which will approach $x=0$ as $b$ gets small, does not coincide with
either the lone energy stationary point near $x=0$ nor the lone variance stationary point near $x=0$.
We therefore conclude that for any $\Gamma\in V_{1,2}$ there exist system/ansatz pairs in which
neither the energy nor the variance is stationary at the $\Gamma$ minimum.

\section{Appendix B}

Here we provide two classes of system that are used in our proof of no size consistency.
First, we define systems of type $B$ in which we target the $E=0$ eigenstate.
Let this system have among its eigenstates the two states
\begin{align}
\hat{H}|B_0\rangle  =  0, \quad
\hat{H}|B_1\rangle  =   |B_1\rangle.
\label{eqn:b_three_eigenstates}
\end{align}
Let the approximate wave function be
\begin{align}
|\Psi\rangle = |B_0\rangle + \alpha |B_1\rangle
\label{eqn:b_wfn}
\end{align}
in which $\alpha$ is a nonzero real number.
We then find that
\begin{align}
\Delta_B = E - 0 = \frac{\alpha^2}{1+\alpha^2}
\quad \mathrm{and} \quad
\sigma^2_B = \frac{\alpha^2}{(1+\alpha^2)^2} > 0.
\label{eqn:b_en_var}
\end{align}

Second, we define systems of type $C$, for which $\Delta=0$ and $\sigma^2>0$
when targeting its $E=1$ eigenstate.
Let this system have among its eigenstates the three states
\begin{align}
\hat{H}|C_0\rangle  =  0, \quad
\hat{H}|C_1\rangle  =   |C_1\rangle, \quad
\hat{H}|C_2\rangle  =  2|C_2\rangle.
\label{eqn:c_three_eigenstates}
\end{align}
Let the approximate wave function be
\begin{align}
|\Psi\rangle = \frac{\beta}{\sqrt{2}} |C_0\rangle + |C_1\rangle + \frac{\beta}{\sqrt{2}} |C_2\rangle,
\label{eqn:second_three_wfn}
\end{align}
in which $\beta$ is a nonzero real number.
We then find that
\begin{align}
\Delta_C = E - 1 = 0
\quad \mathrm{and} \quad
\sigma^2_C = \frac{\beta^2}{1+\beta^2} > 0.
\label{eqn:second_three_ev}
\end{align}


\clearpage

\bibliographystyle{aip}

\end{document}